\documentclass[floatfix,twocolumn,amsmath,amssymb,aps,prb,showpacs,letter]{revtex4}
\usepackage{latexsym,epsfig,bm,times,psfrag,subfigure}

\begin{document}
\title{Spin quantum Hall effect and plateau transitions in multilayer network models}
\author{J. T. Chalker} 
\affiliation{Theoretical Physics, Oxford University, 1, Keble Road, Oxford, OX1 3NP, United Kingdom}
\author{M. Ortu\~no, and A. M. Somoza}
\affiliation{Departamento de F\'isica - CIOyN, Universidad de Murcia,
  Murcia 30.071, Spain}
\date{\today} 
\begin{abstract}
We study the spin quantum Hall effect and transitions between Hall plateaus in quasi 
two-dimensional network models consisting of several coupled layers. Systems exhibiting the
spin quantum Hall effect belong to class C in the symmetry classification for Anderson localisation, and
for network models in this class there is an established mapping between the quantum problem and a classical one 
involving random walks. This mapping permits numerical studies of plateau transitions in much larger
samples than for other symmetry classes, and we use it to examine localisation in systems
consisting of $n$ weakly coupled layers. Standard scaling ideas lead one to expect $n$ distinct plateau transitions, but
in the case of the unitary symmetry class this conclusion has been questioned. 
Focussing on a two-layer model, we demonstrate that
there are two separate plateau transitions, with the same critical properties as in
a single-layer model, even for very weak interlayer coupling.
\end{abstract}


\maketitle

\section{Introduction}
Universality classes for critical behaviour at Anderson transitions are determined by
the dimensionality and the symmetries of the
Hamiltonian.\cite{review}
The best known universality classes are the three Wigner-Dyson 
classes originally identified in
the context of random matrix theory. 
Seven additional symmetry classes for localisation were
recognised\cite{altland-zirnbauer} over a decade ago: they are distinguished from the
Wigner-Dyson classes by having a special point in the energy
spectrum and energy levels that appear in pairs either side of this
point. 
In one of these additional classes, known as
class C, properties of suitably chosen quantum lattice models can be
expressed in terms of observables for a classical model defined on the
same lattice. This mapping was originally discovered\cite{gruzberg} in
the context of the spin quantum Hall effect (SQHE), where it relates a
delocalisation transition in two dimensions to classical
percolation, also in two dimensions, for which many relevant aspects of 
critical behaviour are known
exactly.  

Models in class C arise from the Bogoliubov
de-Gennes Hamiltonian for quasiparticles in a gapless, disordered spin-singlet
superconductor with broken time-reversal symmetry for orbital motion
but negligible Zeeman splitting. 
The special energy in this case is the chemical potential
(which we set to zero) and pairs of levels are related by
particle-hole symmetry, which has profound consequences for
the influence of disorder on quasiparticle eigenstates. 
The quantum to classical mapping provides a framework within which 
quasiparticle properties can be studied in great detail starting from a 
simplified description of a disordered superconductor.

Here we use this approach to study for the SQHE an aspect
of the plateau transition that has resisted detailed investigation in the context of the
conventional integer quantum Hall effect (IQHE) belonging to the unitary symmetry class. 
Specifically, we study plateau transitions in models of $n$ weakly coupled layers. 
For such systems, scaling ideas\cite{khmelnitzkii} and the sigma model 
description\cite{pruisken} lead one to expect $n$ distinct transitions, separating adjacent 
pairs of phases in which the Hall conductance differs by one quantum unit, irrespective of how weakly the layers are coupled. The same scaling ideas have other important consequences. In particular, they suggest
a scenario for the disappearance of the IQHE as magnetic field strength is reduced, in which extended states
responsible for plateau transitions levitate\cite{khmelnitzkii2,laughlin} in energy, and they are input for construction
of the global phase diagram,\cite{global} in which the Hall conductance of adjacent IQHE phases again 
differs by one quantum unit. Alternative types of behaviour have also been proposed, involving 
direct transitions between phases with Hall conductance differing by multiple quantum units,\cite{dissidents}
and there has developed quite an extensive literature on the subject, reviewed in Ref.~\onlinecite{raikh}.
Attempts\cite{lee,wangleewen,hanna,kagalovsky,sorensen,bhatt} to distinguish between these different possibilities using numerical simulations are hampered by the fact
that in the most interesting regimes -- weak magnetic field or weak interlayer coupling -- the localisation
length is never short, making asymptotic behaviour hard to reach. Even in one of the simplest
settings, involving a two-layer IQHE system as a representation of a spin-degenerate Landau level, the existence of two transitions has been inferred only rather indirectly. By contrast, we show in the following for the SQHE
that the mapping to a classical description allows simulation of sufficiently large systems that 
the behaviour expected from scaling and the sigma model can be revealed in considerable detail.

The mapping\cite{gruzberg}  between a single-layer network model for the SQHE and classical percolation
generalises to all models in the same symmetry class that share a set of key features.\cite{beamond} 
A variety of physical quantities of interest for localisation (although not all) can be determined in this way, including
the two-terminal conductance of a finite sample, which will be our main tool.
The generalisation, however, relates the quantum problem to a classical one involving interacting random walks rather
than percolation, so that while much is known analytically about percolation, simulations are required
to study properties of the classical walks. Since these classical simulations are much less computationally intensive
than a direct study of the quantum problem, much larger system sizes are accessible. 
Here we exploit the classical mapping to study the conductance of quasi two-dimensional SQHE systems.
Similar conclusions to the ones we present have been suggested in earlier work,\cite{beamond-thesis}   
but from a much more restricted range of system sizes. 
Our results for quasi two-dimensional systems are complementary to recent work on the
metal-insulator transition in a three-dimensional class C network model,
in which the classical mapping enabled a measurement of
critical exponents with a precision unprecedented for a localisation transition.\cite{OrSo09}

While our focus is on properties of the quantum system, we believe that the classical
walks we study deserve attention in their own right. In particular, it would be interesting if arguments could be found directly for the classical problem, to show that the $n$-layer system has $n$ transitions, and that these are in the
same universality class as classical percolation on the plane.

\section{Model}

We study models in which quasiparticles propagate along the directed links of a lattice and
scatter between links at nodes. Disorder enters the models in the form
of quenched random phase shifts associated with propagation on
links. For class C models, the disorder-averaged (spin) conductance can be 
expressed as an average over
configurations of  interacting classical random walks on the same
directed lattice.\cite{gruzberg,beamond} This relation between
quantum properties and classical walks holds on any graph in which
all nodes have exactly two incoming and two outgoing links.
In the classical problem the connection at each node is a quenched random 
variable having two possible arrangements. Incoming and outgoing links are 
arranged in pairs, and a particle passing through the node
follows the pairing with probability $p$, or switches with probability
$1-p$.
Given a directed graph with the required coordination, 
any choice of classical connections at the
nodes separates paths on the graph into a set of distinct, closed,
mutually avoiding walks. 
Average properties of these walks are calculated from a sum over
all node configurations, weighted according to their probabilities.
 
Here we consider systems formed from several
coupled layers. Each layer is an $L\times L$ square sample
of the L-lattice, shown schematically in Fig.\ 1. 
It is characterised by the node probability $p$ and in isolation 
has a plateau transition at $p=1/2$. We construct $n$-layer models
by stacking $n$ copies of this lattice in register, with independent disorder
realisations in each layer. For $n=2$ we couple the layers
using a second set of nodes, located at the mid-points of the links in each layer.
This set of nodes is characterised by the probability $p_1$ of switching layers, and the
model is symmetric under $p_1 \to 1-p_1$.
At $p_1=0$, and also at $p_1=1$, the system separates into independent layers, both
with a plateau transition at $p=1/2$; we are concerned with behaviour as a function of $p$
for $0< p_1 < 1$, and in particular whether two coupled layers exhibit two transitions
at separate values of $p$. We also examine, though in less detail, a three-layer system,
considering only a parameterless form of interlayer coupling. This is constructed by allowing
at the mid-points of the links in each layer all six permutations of trajectories between layers, with
equal probability.

\begin{figure}[htb]
\includegraphics[width=.22\textwidth]{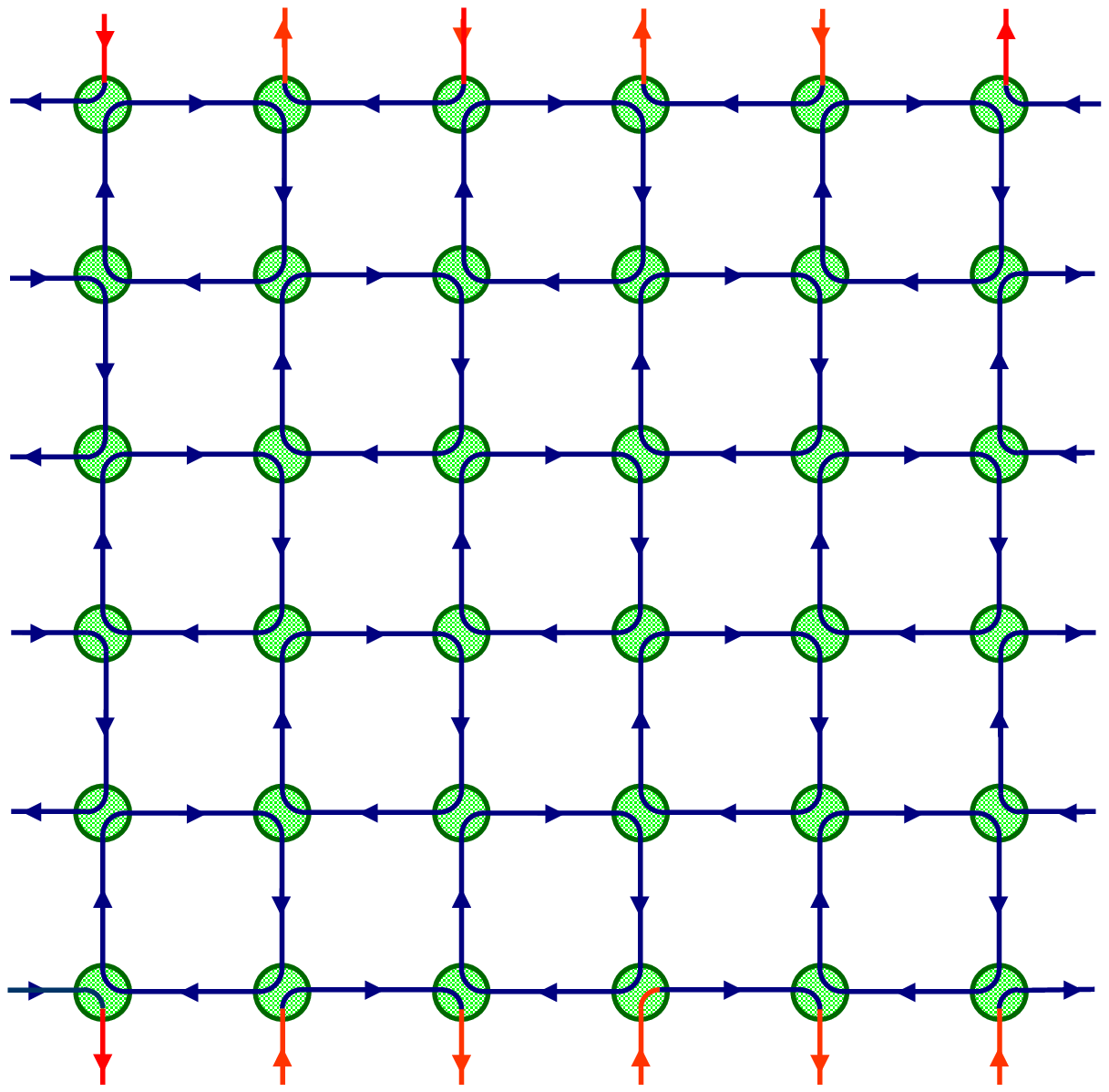}
\hskip .1cm\includegraphics[width=.22\textwidth]{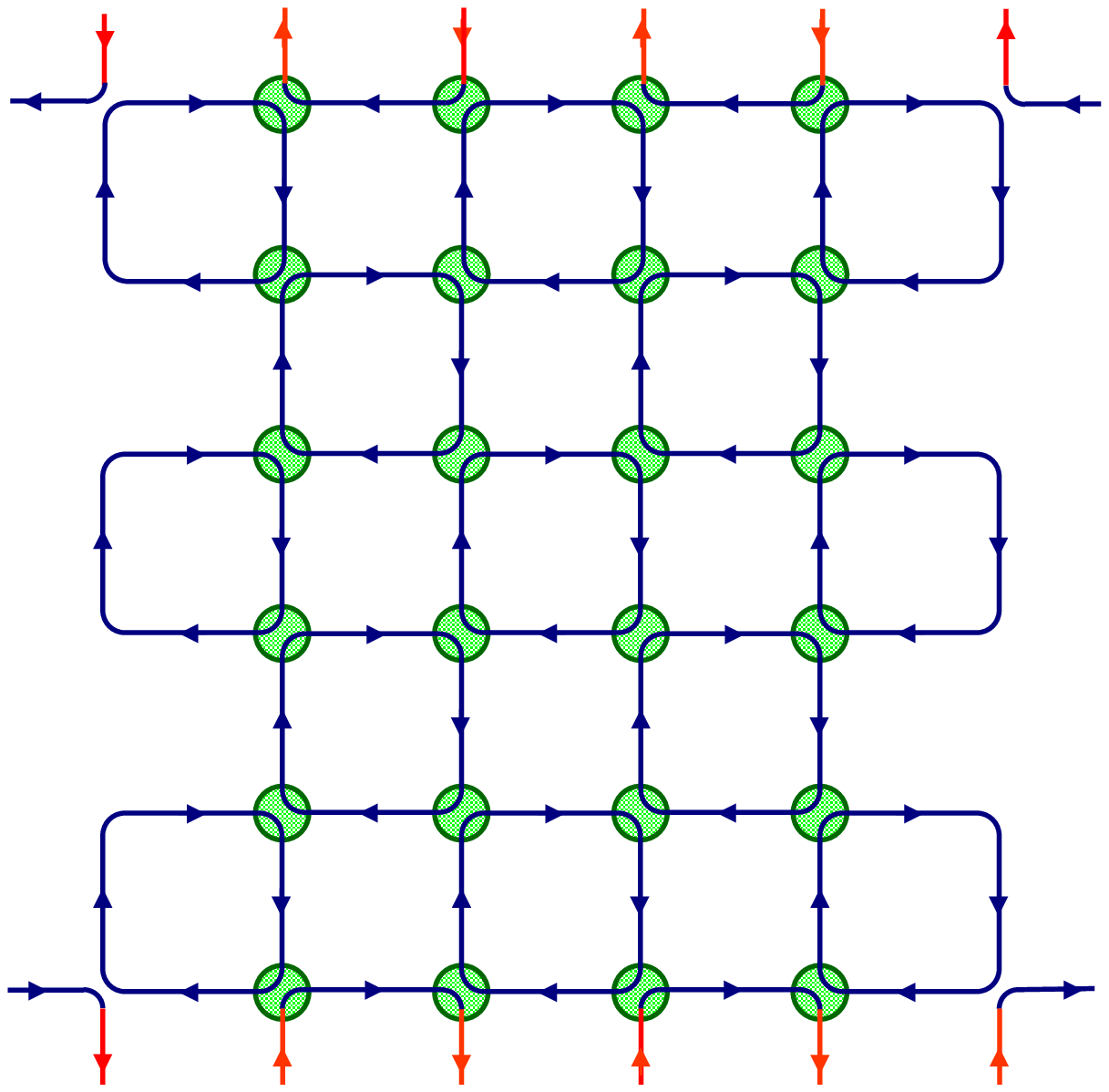}
\caption{(Color online) The L-lattice: circles represent nodes at which links are connected as
indicated with probability $p$, and in the opposite sense with probability $1-p$. Left and right 
panels indicate the boundary conditions applied to obtain the quantities we refer to respectively as 
the longitudinal conductance and Hall conductance.
In both cases, current leads are attached to the left and right edges. For calculation of the longitudinal conductance,
periodic boundary conditions are applied to the top and bottom edges to form a cylinder. For the Hall conductance, reflecting boundary conditions 
are applied to the top and bottom edges.}
\label{fig1}
\end{figure}

We calculate the two-terminal conductance between opposite, open faces as the average number of classical 
trajectories connecting these two faces. We use
two different types of boundary condition in the other direction, as illustrated in Fig.\ 1.
With periodic boundary conditions, so that the sample is a cylinder, the two-terminal
conductance of a system described by a constant local conductivity tensor would be simply
the longitudinal conductivity. For that reason we call the average conductance in this geometry
the {\it longitudinal conductance}. 
Alternatively, using reflecting boundary conditions, the two-terminal conductance 
within a Hall plateau is determined by the number of edge states.
We therefore call the average conductance in this geometry the {\it Hall conductance},
even though its value between Hall plateaus  depends on both components of the conductivity tensor.
Within the framework of the quantum-to-classical mapping we use,
the disorder-averaged spin conductance $G(p,L)$ of the quantum system is given (in units of $\hbar/4\pi$) by
the average of the number of classical paths from a
specified open face to the other. 

Our simulations use system sizes $L$ of between 500 and 5000 lattice spacings. 
For the largest system, we average up to $10^6$ disorder realisations. Earlier
work,\cite{beamond-thesis}  was limited to $L\leq 80$.

\section{Conductance and spin quantum Hall transitions}

\begin{figure}[h]
\includegraphics[width=.48\textwidth]{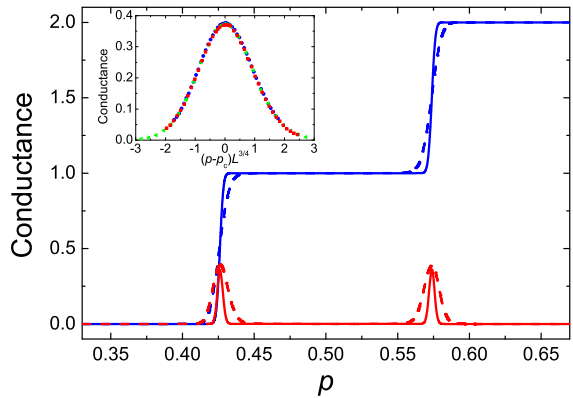}
\caption{(Color online) Conductance of the two-layer system at $p_1=1/2$  as a function of $p$ 
for system sizes $L=1000$ (dashed lines) and 4000 (continuous lines). Curves with steps (blue) 
represent the Hall conductance, obtained with reflecting boundary conditions. 
Curves with peaks (red) are the longitudinal conductance, obtained with a cylindrical sample.
The inset shows the scaling of the longitudinal conductance near the first peak as a function of 
$(p-p_{\rm c})L^{3/4}$ for $L=3000$, 4000 and 5000.
}
\label{fig2}
\end{figure}

We first study the two-layer system at interlayer coupling  $p_1=1/2$.
In Fig.\ \ref{fig2} we show the behaviour of the longitudinal and transverse conductances
as a function of the intralayer parameter $p$ for two system sizes.
Two transitions are apparent, at $p\simeq 0.43$ and $p\simeq 0.57$, separating three
phases characterised by quantised Hall conductances of 0, 1 and 2 units.
The accurate quantisation of the Hall conductance in the central phase is striking. 
In terms of classical walks, it arises because almost all realisations that contribute
to the average have exactly one extended trajectory at each reflecting edge.
Other trajectories in this phase are typically much shorter than sample size, since the longitudinal 
conductance is very small.

These transitions in the two-layer system are expected to
have the same critical behaviour as for a single layer, which maps to that of the percolation transition.
As a first test, we plot in the inset to Fig.\ \ref{fig2} the longitudinal conductance
near the peak at $p\simeq0.43$ as a function of $(p-p_{\rm c})L^{3/4}$ for three system sizes
$L=3000$ (blue dots), 4000 (green triangles) and 5000 (red squares). The good overlap
of the data for different sizes shows that the widths  of
the peaks in longitudinal conductance scale as $L^{-1/\nu}$, 
where $\nu=4/3$ is the correlation length exponent for percolation. 
The width of the steps in Hall conductance follows a similar behaviour.

\begin{figure}[h]
\includegraphics[width=.48\textwidth]{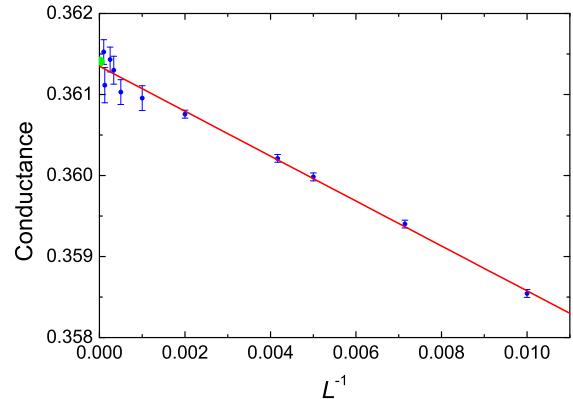}
\caption{(Color online) Conductance of a single-layer system at criticality, $p=1/2$, versus $L^{-1}$.
The large dot on the vertical axis corresponds to the exact value.
}
\label{fig3}
\end{figure}

\begin{figure}[h]
\includegraphics[width=.48\textwidth]{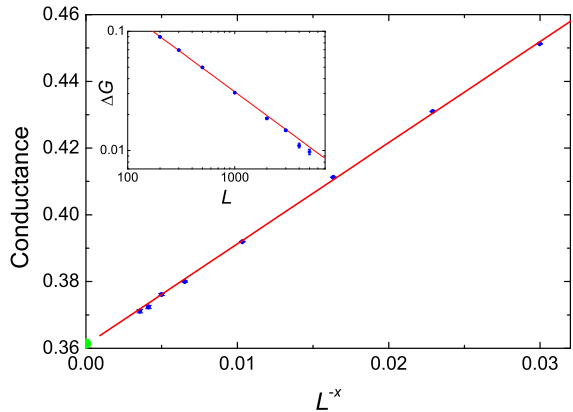}
\caption{(Color online) Maximum value of the conductance of a two-layer system as a function of $L^{-0.662}$.
The large dot on the vertical axis corresponds to the exact value for percolation. Inset: maximum conductance,
with respect to the percolation value, versus $L$ on a double-logarithmic scale.
}
\label{fig4}
\end{figure}

As a second and more precise check of universality, we
examine the peak value of the longitudinal conductance. In a single layer this can be deduced from the
crossing probabilities for percolation clusters in an annulus.\cite{gruzberg,cardy1} We obtain values for
these from Ref.~\onlinecite{cardy02}:
the probability of a single crossing in a square sample with cylindrical boundary conditions 
is $0.357369\ldots$ and the probability of two crossings 
is $0.002018\ldots$, while higher order crossings are neglegible at our precision. So
the exact value of the critical conductance in this geometry from percolation theory is $0.361404\ldots$. 
To test our numerical approach, we first calculate the maximum value of the longitudinal conductance for a single layer
using a sequence of system sizes at the exact critical probability for percolation, $p=1/2$. 
We find that deviations of these maxima from the large system limit decrease 
roughly proportionally to $L^{-1}$, so we extrapolate to infinite size 
by plotting maxima as a function of $L^{-1}$. We show the result in Fig.\ \ref{fig3}. 
A linear fit yields the intercept $0.36135\pm 0.00010$, confirming the reliability of the approach.
We also obtain the maximum conductance for a bilayer system, but  this time we do not know the exact critical 
probability for percolation so we have to calculate the conductance for several probabilities in the critical region
and fit the results with a Gaussian to extract the maximum value at each system size. The deviations of this maximum value 
from the large system limit do not scale as either $L^{-1}$ or $L^{-3/4}$ for the range of system sizes considered.
To extract a limiting value of the critical conductance for the bilayer system, we therefore 
first plot the difference between the bilayer maximum conductances and the critical conductance given by percolation
theory, as a function of $L$
on  a double-logarithmic scale. The result is shown in the inset to Fig.\ \ref{fig4}: the data lie close to a straight line
with slope $x=0.662$. We then plot in the main part of Fig.\ \ref{fig4} the conductance as a function of 
$L^{-x}$. This figure shows that the conductance extrapolates to a value close to the one from percolation theory, and we obtain
$0.3613 \pm 0.0002$. The uncertainty is 
higher than for a single layer because the critical value of $p$ is not known exactly
and because the system sizes employed are necessarily smaller.

We have also studied the conductance for the three-layer model described above. Results are
shown in Fig.\ 5. As expected, three distinct transitions separate four phases, and adjacent phases
have Hall conductances differing by a single unit; the critical points are at $p\simeq 0.37$, $0.5$ and $0.63$.

\begin{figure}[h]
\includegraphics[width=.48\textwidth]{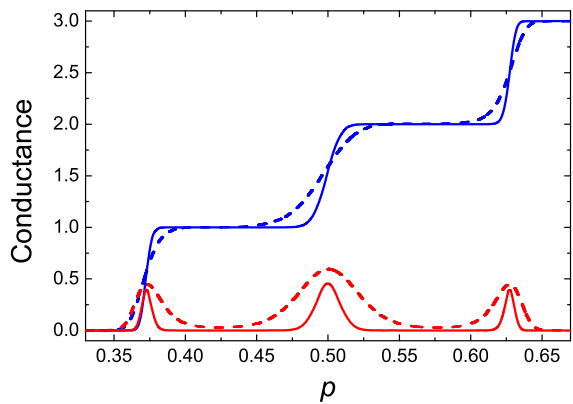}
\caption{(Color online) Conductance as a function of $p$ for the three-layer system defined in the text, with system
sizes $L=1000$ (dashed lines) and 4000 (continuous lines). 
Curves with steps (blue) 
represent the Hall conductance, obtained with reflecting boundary conditions. 
Curves with peaks (red) are the longitudinal conductance, obtained with a cylindrical sample.
}
\label{fig5}
\end{figure}

\section{Phase diagram for a two layer system}

\begin{figure}[h]
\includegraphics[width=.48\textwidth]{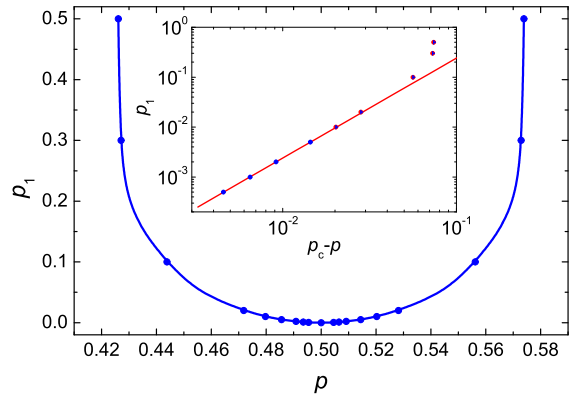}
\caption{(Color online) Phase boundaries for spin quantum Hall states
in a two-layer system, as a function of the intralayer parameter $p$ and the interlayer parameter $p_1$.
Points are from extrapolations described in the main text; the continuous curve is a guide to the eye.
Inset: same data on a double logarithmic scale.
}
\label{fig6}
\end{figure}

The results presented in the previous section are for maximal coupling between layers, and it is
interesting to examine how behaviour changes as this coupling is reduced. In particular, a key question is whether
the degenerate transition occurring at $p=1/2$ for $n$ uncoupled layers is split into $n$ distinct transitions
by any non-zero interlayer coupling, or whether a step in Hall conductance of more than one unit persists for
small couplings. We focus on properties of the two-layer system as a function of the interlayer coupling $p_1$,
and calculate the longitudinal conductance rather than Hall conductance because it is
easier to locate the transition probabilities using this quantity.
For all system sizes and values of $p_1$ at which there are two clear peaks in the 
longitudinal conductance as a function of $p$, we determine peak positions $p_{\rm c}$ by fitting the conductance of one of them to a Gaussian in $p-p_{\rm c}$. At each value of  $p_1$ we make an extrapolation of these peak positions to infinite system size, linearly in $L^{-\alpha}$. We find empirically that $\alpha=2$ is the best fitting exponent in all cases considered,
but the results do not depend much on this value. We consider sizes $L$ from 500 to 5000. 
In large systems it is possible to identify two distinct transitions even at very weak interlayer coupling. For example, with $p_1=5\cdot 10^{-4}$ we can distinguish two peaks for sizes greater than 1000, and for $p_1=10^{-3}$ we can discriminate for sizes greater than $500$. The extrapolated values of the critical probability for the transitions are shown in Fig.\ \ref{fig6}. 

To gain further insight, we analyse in detail the shape of the phase boundary 
at small interlayer coupling. In this regime $p_{\rm c}$ is close to $1/2$ and if two distinct transitions
persist for all non-zero $p_1$, we expect a relation of the form
\begin{equation}\label{gamma}
p_1\propto |p_{\rm c}-1/2|^\gamma .
\end{equation}
In the inset to Fig.\ 4 we show $p_1$ versus $p_{\rm c}-1/2$,  on a double
logarithmic scale.
The straight line is a linear fit to the
four points with smallest $p_1$. Its slope is $\gamma=2.004\pm 0.002$.
The fact that the phase boundary is accurately described by Eq.~(\ref{gamma}) is good evidence
for the correctness of the quantum Hall scaling flow diagram of Ref.~\onlinecite{khmelnitzkii}
when applied to the SQHE. It is also evidence that in the system studied there is no direct transition between phases
with Hall conductance differing by more than one unit.

The value $\gamma=2$ can be understood by the following argument. Consider first a single-layer system, and
let $\xi(p)$ be the correlation length for classical walks: the typical diameter of the largest closed loops.
These loops are the hulls of percolation clusters, and for large $\xi(p)$ their arc-length varies as $\xi(p)^{d_{\rm h}}$ with $d_{\rm h}=7/4$. Moreover, from the mapping\cite{gruzberg} for a single layer to classical percolation, we know that $\xi(p)$
diverges as $\xi(p) \propto |p-1/2|^{-\nu}$ with $\nu=4/3$ when $p$ approaches $1/2$.
Next examine the probability in a two-layer system that a pair of such loops, one from each layer, are
coupled. This probability is expected to be of order one for $p=p_{\rm c}$. It is made up of the product of three factors: the length $\xi(p)^{d_{\rm h}}$ of one loop, the density $\xi(p)^{d_{\rm h}-2}$ of the other loop and the probability $p_1$ that a given pair of links in equivalent positions in the two layers are coupled. We therefore expect $p_1 \propto \xi(p_{\rm c})^{2-2d_{\rm h}}$ and hence $p_1 \propto |p-1/2|^{2\nu(d_{\rm h} - 1)} =  |p-1/2|^2$. It is interesting to note
the difference between this result for the SQHE and the equivalent one for a two-layer IQHE system, in which\cite{dohmen}
(taking over the notation we have defined for the SQHE) $p_1 \propto |p_{\rm c} - 1/2|^{\nu}$. In the language of the quantum localisation problem, this difference arises because the density of states vanishes at the mobility edge for the SQHE but is finite for the IQHE.

This work was supported by EPSRC Grant No. EP/D050952/1, by DGI
Grant No. FIS2009-13483, and by Fundacion Seneca, Grant No. 08832/PI/08.

\end{document}